\documentclass{PoS}
\usepackage{url}
\usepackage{microtype}

\title{Electron-Ion Physics with the LHeC}

\ShortTitle{Electron-Ion Physics with the LHeC}

\author{\speaker{Ilkka Helenius}$^{\mbox{}\,a}$, Hannu Paukkunen$^{b,c}$ and N\'{e}stor Armesto$^{d}$ for the LHeC Study Group\\
\llap{$^a$}Department of Astronomy and Theoretical Physics, Lund University, S\"{o}lvegatan 14A,\\ SE-223 62 Lund, Sweden\\
\llap{$^b$}Department of Physics, University of Jyvaskyla, P.O. Box 35, \\FI-40014 University of Jyvaskyla, Finland\\
\llap{$^c$}Helsinki Institute of Physics, P.O. Box 64, FI-00014 University of Helsinki, Finland\\
\llap{$^d$}Departamento de F\'{\i}sica de Part\'{\i}culas and IGFAE, Universidade de Santiago de Compostela,\\E-15706 Santiago de Compostela, Galicia-Spain\\
        E-mail: \email{ilkka.helenius@thep.lu.se}, \email{hannu.t.paukkunen@jyu.fi}, \email{nestor.armesto@usc.es}}

\abstract{The Large Hadron Electron Collider (LHeC) project is the proposal to use the existing LHC proton/ion beams and construct a new electron beam line to perform high-energy electron-proton/ion collisions. In this talk, we consider some of the physics topics that could be studied in the electron-ion mode. In particular, we estimate how much the current nuclear parton distribution fits could be improved with the deeply inelastic scattering measurements at the LHeC by including pseudodata into a global analysis. In addition, we discuss briefly other topics that would help to better understand some aspects of heavy-ion collisions, namely small-$x$ physics and hadron production with a nuclear target.}

\FullConference{XXIII International Workshop on Deep-Inelastic Scattering\\
		 27 April - May 1 2015\\
		 Dallas, Texas}

\begin{document}

\section{Introduction}

Experimentally and theoretically, the cleanest environment to measure parton distribution functions (PDFs) describing the partonic structure of nucleons is the Deeply Inelastic Scattering (DIS) process where the target hadron is probed with a lepton beam. There has been a long history of such experiments starting from fixed target electron-proton collisions at SLAC culminating to the HERA collider which collided a $27.5 \ \mathrm{GeV}$ electron/positron beam with a $920 \ \mathrm{GeV}$ proton beam. For the Large Hadron Electron Collider (LHeC) \cite{AbelleiraFernandez:2012cc}, the plan is to use the proton/ion beams from the LHC and collide it with an electron beam from a new accelerator providing $50-100\ \mathrm{GeV}$ electrons. In the future the new electron beam could also be used in conjunction with the hadron beams from the projected Future Circular Collider (FCC) to further increase the collision energy.

The DIS measurements at HERA provide precise constraints for the proton PDF analyses down to  $x\sim10^{-5}$, where $x$ is the momentum fraction of the parton with respect to momentum of the hadron. However, for nuclear PDFs (nPDFs) there are no collider DIS data available. As the kinematic reach of the fixed target experiments is limited to $x \gtrsim 0.005$, the current nPDF fits \cite{Eskola:2009uj, Kovarik:2015cma, deFlorian:2011fp, Hirai:2007sx} are roughly unconstrained below this $x$-value. This means that the baseline for  heavy-ion physics at the LHC is not under good control at the moment. The data from proton-lead collisions at the LHC provide some further constraints for the nPDFs but even these new data do not provide sensitivity to the small-$x$ region. Here we will show,  by performing a new global analysis using pseudodata with realistic uncertainties, how the LHeC measurements would improve the precision of the nPDF analyses.

In addition to the nPDF studies, the small-$x$ region is interesting also because a breakdown of the linear DGLAP evolution is expected when high-enough parton density is reached. This and other physics opportunities at the LHeC in electron-ion mode are briefly discussed as well.

\section{Nuclear PDFs}

The DIS cross sections can be written in terms of structure functions $F_j(x,Q^2)$ ($j=1,2,3$) describing the structure of the target hadrons. In the collinear factorization  approach the observed modifications of the structure functions with nuclear target are absorbed into process independent nPDFs $f_i^A(x,Q^2)$. These nPDFs can be extracted from data by performing a global QCD analysis. There are two different ways to do this: one can either parametrize the $f_i^A(x,Q^2)$ directly (e.g. nCTEQ \cite{Kovarik:2015cma}) or one can take some proton PDF set as a baseline and parametrize the nuclear modification $R_i^A(x,Q^2)$ (e.g. EPS09 \cite{Eskola:2009uj}) defined as
\begin{equation}
R_i^A(x,Q^2) = \frac{f_i^A(x,Q^2)}{f_i(x,Q^2)},
\end{equation}
where $f_i(x,Q^2)$ is the free proton PDF.\footnote{A set of proton PDFs is required also in the former option  in order to compute the nuclear modifications in $F_j(x,Q^2)$. Thus, there is no fundamental difference between the two.} In both cases the parameters are fixed by evolving the parametrized nPDFs by the DGLAP equations from an initial scale $Q_0$ and comparing the calculated cross sections to the available data. The accuracy of the globally analyzed nPDFs are thus dictated by the precision and kinematic reach of the included data.

\begin{figure}[htb]
\begin{minipage}[t]{0.49\textwidth}
\centering
\includegraphics[width=0.972\textwidth]{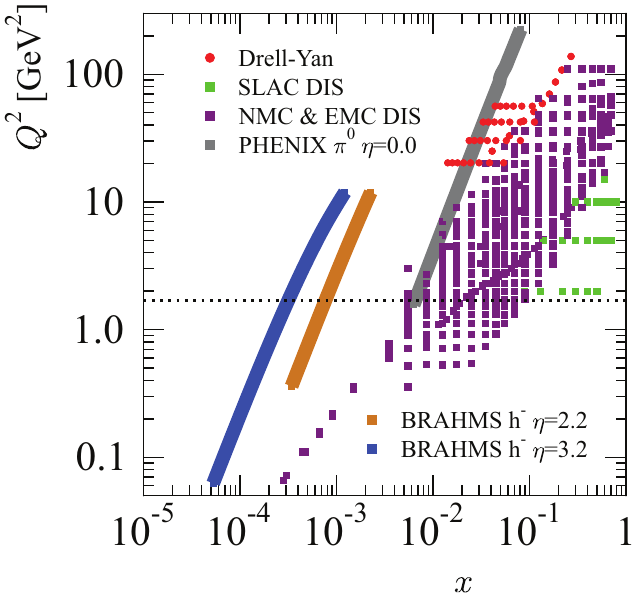}
\caption{The kinematic coverage of the different data sets used in EPS09 global analysis. The dashed line shows the applied lower cut in $Q^2$ in the analysis. Figure from Ref.~\cite{Eskola:2009uj}.}
\label{fig:kinReachEPS09}
\end{minipage}%
\hspace{0.02\linewidth}
\begin{minipage}[t]{0.49\textwidth}
\centering
\includegraphics[width=\textwidth]{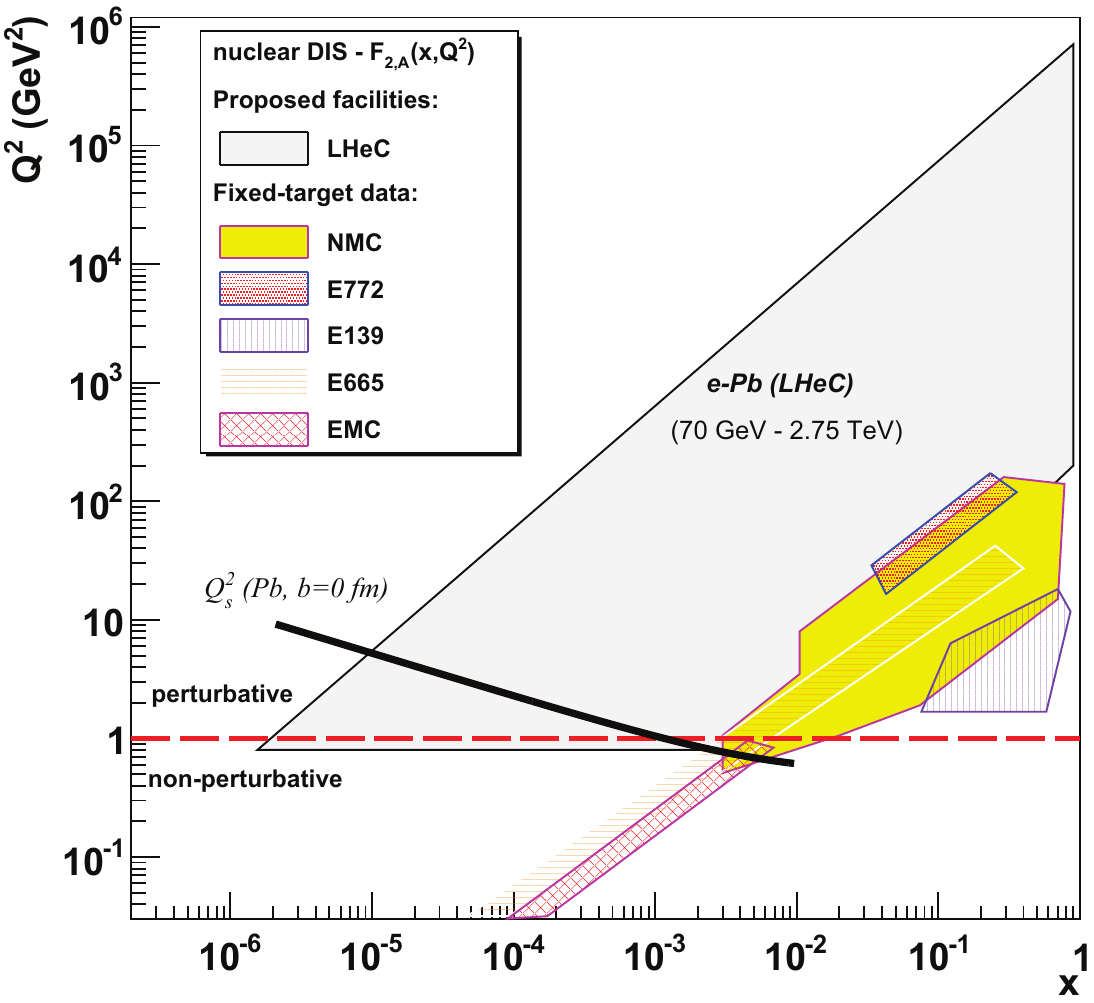}
\caption{The expected kinematic coverage of the LHeC nuclear DIS data compared to existing fixed-target nuclear DIS data. Figure from Ref.~\cite{AbelleiraFernandez:2012cc}.}
\label{fig:kinReachLHeC}
\end{minipage}
\end{figure}

The current nPDF fits are mainly constrained by fixed-target DIS and Drell-Yan dilepton production data. Additional constraints are obtained from inclusive pion production in d+Au collisions at RHIC (EPS09, nCTEQ and DSSZ \cite{deFlorian:2011fp}) and neutrino DIS (DSSZ). Figure \ref{fig:kinReachEPS09} shows the kinematic reach of the data in $(x,Q^2)$-plane used in the EPS09 analysis\footnote{BRAHMS data have not been included in any analysis.} and figure \ref{fig:kinReachLHeC} shows the expected coverage of the DIS measurement at the LHeC. The increase of the kinematic range is more than three orders of magnitude both in $x$ and $Q^2$ which would be a huge improvement to the prevailing situation. 

{\phantom{h}}

\begin{figure}[htbp]
\centering
\includegraphics[trim = 10pt 545pt 80pt 55pt, clip, width=\textwidth]{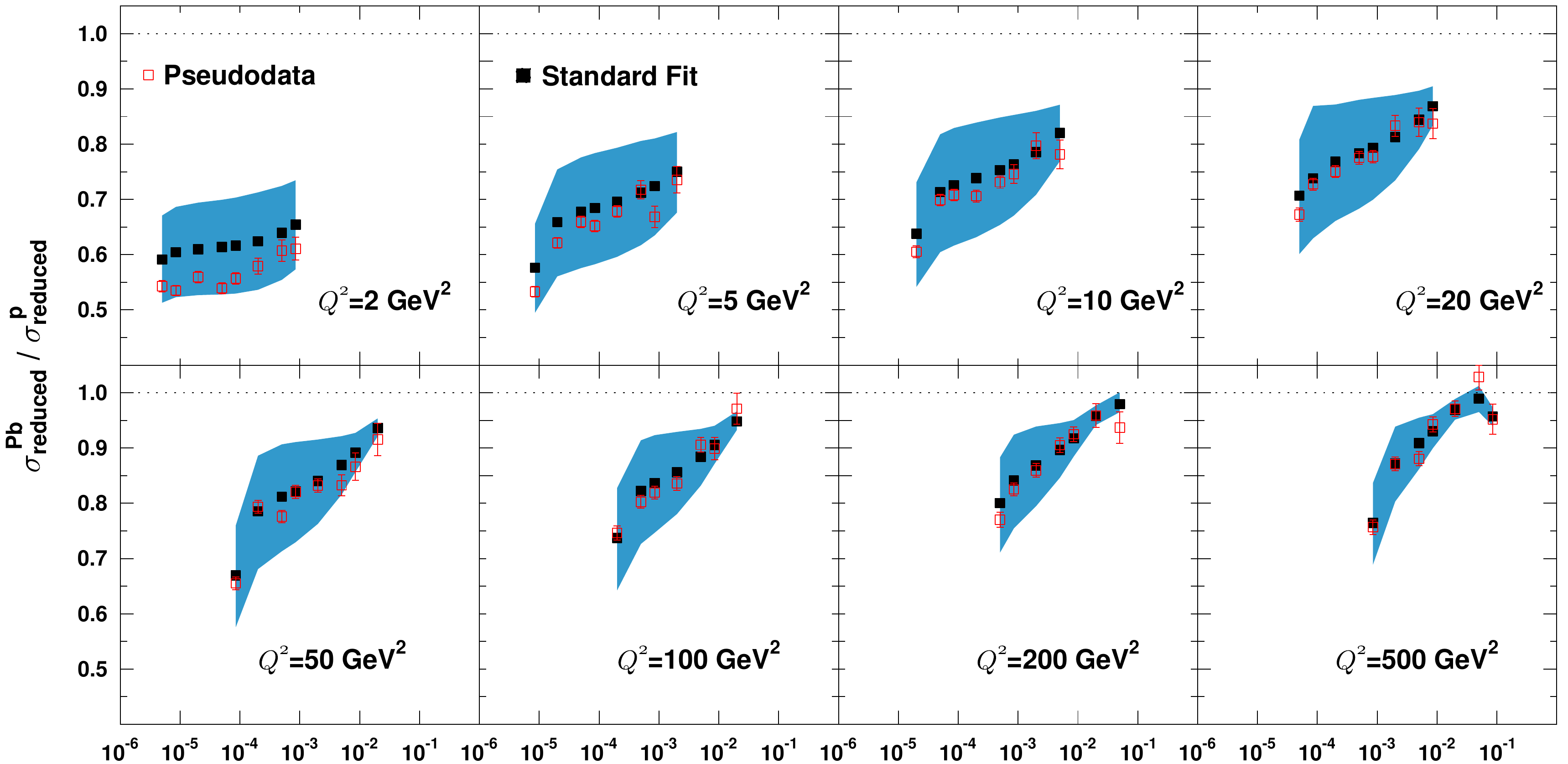}
\caption{The ratio between the reduced neutral current DIS cross section with lead and proton target using $60\,\mathrm{GeV}$ electron beam. The generated pseudodata are shown in red together with the estimated uncertainties and baseline-fit predictions in black. The blue bands show the baseline fit uncertainty. Figure from Ref.~\cite{Paukkunen:2014slides}.}
\label{fig:DataNoLHeC}
\end{figure}

\begin{figure}[htbp]
\centering
\includegraphics[trim = 10pt 545pt 80pt 55pt, clip, width=\textwidth]{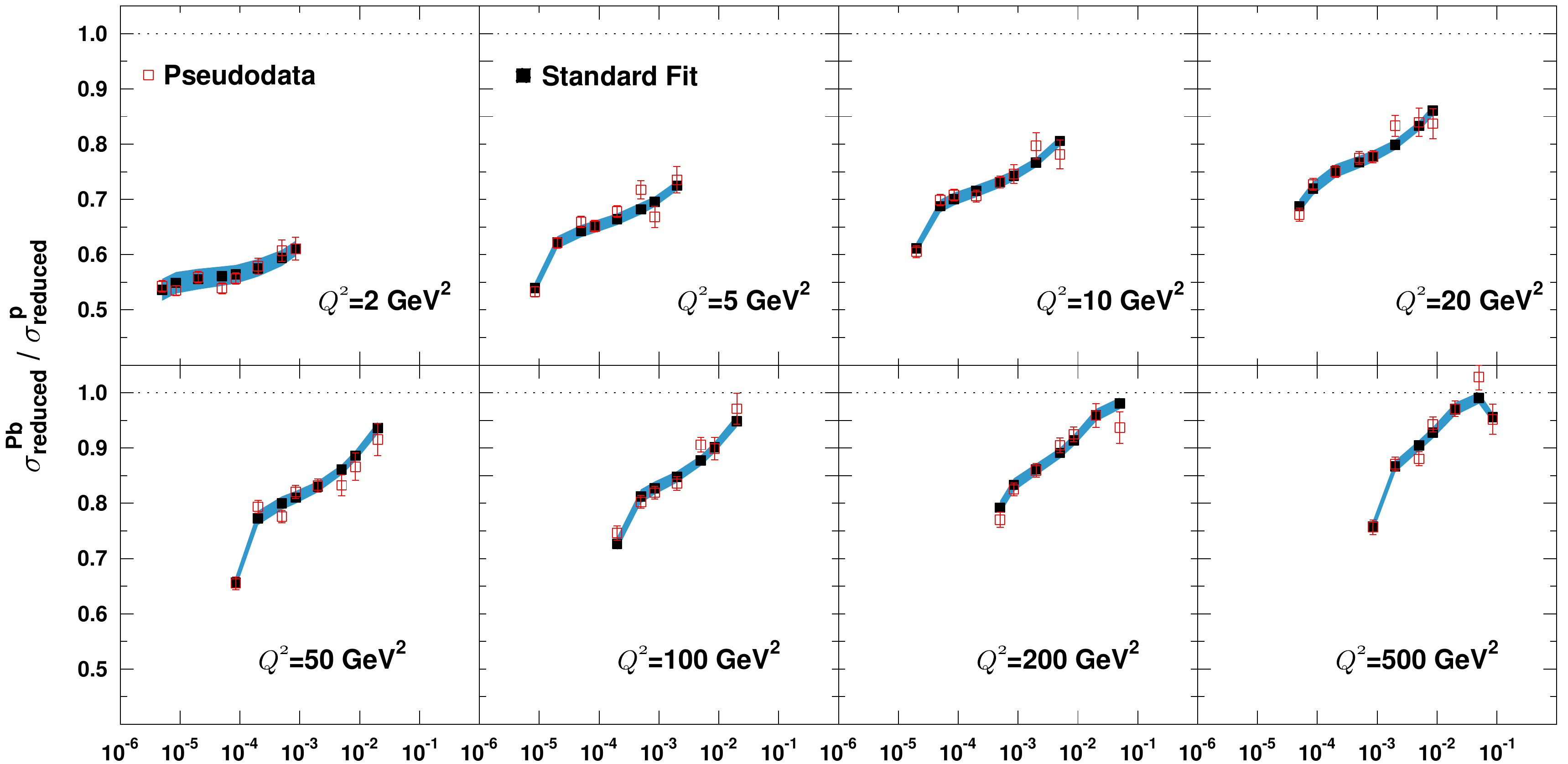}
\caption{Same as figure \protect\ref{fig:DataNoLHeC} but after including the pseudodata into the fit. Figure from Ref.~\cite{Paukkunen:2014slides}.}
\label{fig:DataWithLHeC}
\end{figure}

The potential of the LHeC DIS data can be quantified by generating a set of pseudodata and performing a re-analysis using this set on top of the existing data, and comparing the results  with the previous fits. The pseudodata are generated according to the theoretical expectations including realistic estimate for experimental uncertainties. Here samples of neutral current DIS reduced cross section $\sigma_{\rm reduced}$ were generated within $10^{-5}<x<1$ and $2<Q^2<10^{5}\,\mathrm{GeV^2}$, for more details see Ref.~\cite{Paukkunen:2014slides}. The ratios of $\sigma_{\rm reduced}$ between the lead  and proton targets are shown in figure~\ref{fig:DataNoLHeC} for different $Q^2$ values up to $500\,\mathrm{GeV^2}$ as a function of $x$, both for the generated pseudodata and for the theoretical expectation from a baseline fit (an EPS09-style fit) including the uncertainties. Figure \ref{fig:DataWithLHeC} shows the same data but now compared to the theoretical calculations after including the pseudodata into the fit. The inclusion of this pseudodata set clearly improves the precision of the theoretical result.

\begin{figure}[htb]
\centering
\includegraphics[width=\textwidth]{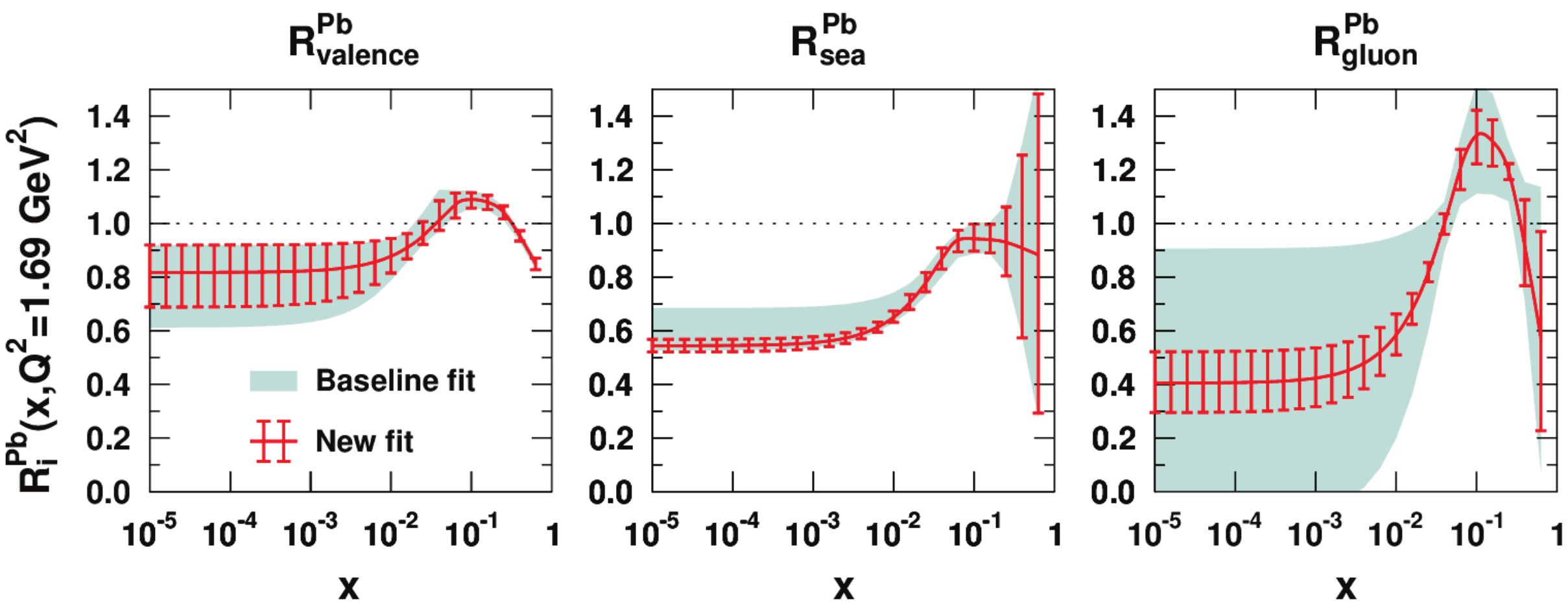}
\caption{The nuclear modification of the valence quarks (left), sea quarks (middle) and gluons (right) and their uncertainties from the baseline fit (blue band) and from the fit including the LHeC pseudodata (red bars). Figure from Ref.~\cite{Paukkunen:2014slides}.}
\label{fig:nPDFsUncert}
\end{figure}

To see the effect of LHeC pseudodata on the nPDF uncertainties we can compare these uncertainties before and after the inclusion of this dataset. This comparison is shown in figure \ref{fig:nPDFsUncert} for valence quarks, sea quarks and gluons at the parametrization scale $Q_0=1.69\,\mathrm{GeV^2}$. Due to the lack of constraints at small values of $x$, especially the gluon uncertainties are large in this region in the baseline fit. However, when the LHeC pseudodata are included these uncertainties are significantly reduced. Also the precision of the sea quark nuclear modification is greatly improved at $x<0.01$. For the valence quarks the effect is not that significant as the large $x$-region is already well constrained and as the quark distributions at small-$x$  are dominated by the sea quark contribution. 
However, one should keep in mind that the fit funcions in these fits are rather restricted (those of EPS09) and do not allow much freedom in the small-$x$ region. In principle the current data would allow a very different small-$x$ behaviour but, as there are theoretical arguments why one expects to have the shadowing effect, it was built into the EPS09 fit functions (used in both fits discussed here). When there will be data in this region one should allow an adequate amount of freedom in the fit function to reduce the bias caused by too rigid fit functions. This extra freedom affects the shape of the small-$x$ behaviour and can also increase the uncertainties. For preliminary studies about this topic see Ref.~\cite{Paukkunen:2015slides}.

Another future improvement would be the inclusion of charged current DIS that would help to constrain the flavor dependence of nuclear modifications. Currently there no data that would be sensitive to flavor separation\footnote{The use of neutrino-nucleus DIS data for this purpose is complicated by the facts that already the \emph{free proton} fits use these data and that the data are available only as absolute cross sections (not ratios).} so usually the nuclear modifications are taken to be flavor independent. The new data from $W^{\pm}$ production in proton-lead collisions from CMS \cite{Khachatryan:2015hha} may provide some handle for the flavor separation but the LHeC charged current DIS would be able to constrain this more decisively \cite{PaukkusenPuuhat}.

\section{Other electron-ion physics opportunities}

\subsection{Small-$x$ physics at the LHeC}

\begin{figure}[htbp]
\centering
\includegraphics[width=0.6\textwidth]{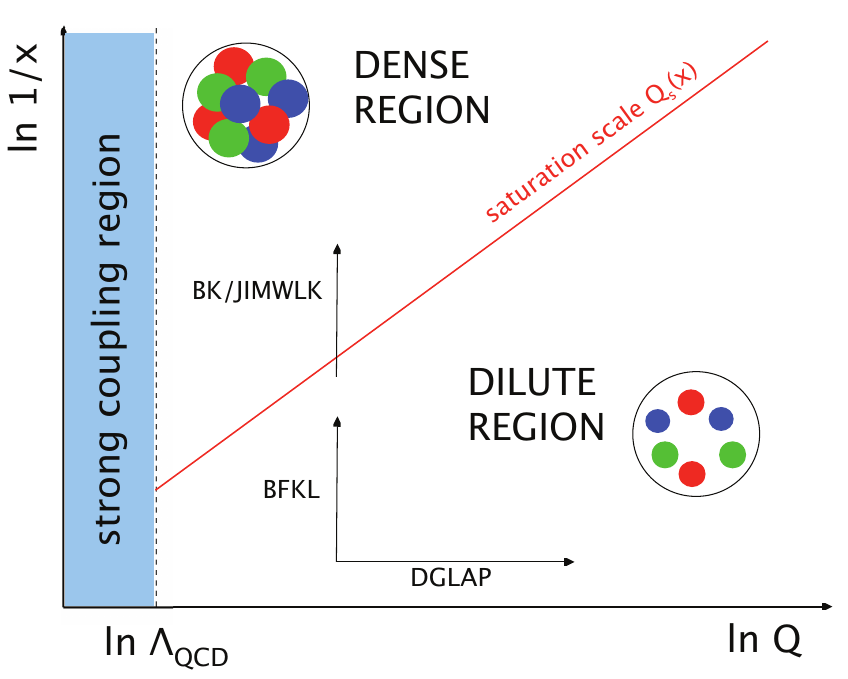}
\caption{A schematic diagram of the behaviour of the expected saturation scale in $Q^2$ and $x$. Figure from Ref.~\cite{AbelleiraFernandez:2012cc}.}
\label{fig:smallx}
\end{figure}

As the LHeC is designed to be capable of probing $x$ values down to $10^{-6}$, it would be an ideal tool to study saturation effects. The linear QCD evolution is expected to break down when the partonic density reaches the point where individual nucleons start to overlap with each other as illustrated in figure~\ref{fig:smallx}. This phenomenon is typically referred to as a saturation effect, see a recent overview in Ref.~\cite{Albacete:2014fwa} for more detailed discussions. The onset of this effect is characterized by the saturation scale $Q_s$ which is expected to roughly behave as $A^{1/3}\,x^{-0.3}$. Due to the nuclear mass number dependence the saturation effects are more pronounced for heavy-ion targets. So far there has not been clear evidence for these effects at the LHC but the clean measurement and the broad kinematic reach would make electron-ion collisions at the LHeC a very promising environment for searching  these effects as indicated in figure \ref{fig:kinReachLHeC} that shows the expected $Q_s(x)$ with a Pb target.

\subsection{Hadronization in nuclear medium}

As  hadron production in proton-lead collisions at the LHC has turned out to be more complicated than anticipated \cite{Khachatryan:2015xaa, ATLAS:2014cza}, it would be important to study a more cleaner environment in order to benchmark the hadron production mechanism on nuclear targets. For low energy hadrons,  hadronization can happen inside the nucleus, possibly leading to absorption before the hadron is formed. For higher energy hadrons, hadronization is likely to happen outside the target but still some amount of the partonic evolution can take place inside the nucleus. An attempt has been made to quantify these kind of modifications by performing a global analysis of nuclear fragmentation functions in Ref.~\cite{Sassot:2009sh}, but more detailed measurements are necessary to confirm these effects.

\section{Conclusions and Outlook}

Electron-ion collisions at the LHeC, where an unprecedented energy will be reached, have a huge potential to reduce the baseline uncertainties for heavy-ion physics. Perhaps this is most apparent in the case of nuclear PDFs. High-precision electron-ion collider data would bring the nPDF analysis to a similar level of accuracy as free proton PDFs are at the moment. It would also help to relax some assumptions in the fits, e.g. flavor dependence, that are currently necessary as they are not well restricted by the data. One should also study how the form of the fit function affects the uncertainty estimates to be able to quantify the realistic improvements that the LHeC would provide. In addition to the nPDFs, there are also other phenomena that are currently under discussion such as saturation at small values of $x$ and hadronization in nuclear environment. Both could be accurately studied with electron-ion collisions at the LHeC.

\section*{Acknowledgments}

I.H. is supported by the MCnetITN FP7 Marie Curie Initial Training Network, contract PITN-GA-2012-315877.
N.A. is supported by  People Programme (Marie Curie
Actions) of the EU Seventh Framework Programme FP7/2007-2013/ under
REA grant agreement \#318921, the European Research Council
grant HotLHC ERC-2011-StG-279579, Xunta de Galicia (Conseller\'{\i}a de Educaci\'on and Conseller\'{\i}a de Innovaci\'on
e Industria - Programa Incite), the Spanish Consolider-Ingenio 2010 Programme
CPAN and FEDER.

\mbox{}


\begin{thebibliography}{99}

\bibitem{AbelleiraFernandez:2012cc}
  J.~L.~Abelleira Fernandez {\it et al.} [LHeC Study Group Collaboration],
  \emph{A Large Hadron Electron Collider at CERN: Report on the Physics and Design Concepts for Machine and Detector},
  \emph{J.\ Phys.\ G} {\bf 39} (2012) 075001
  [arXiv:1206.2913 [physics.acc-ph]].

\bibitem{Eskola:2009uj}
  K.~J.~Eskola, H.~Paukkunen and C.~A.~Salgado,
  \emph{EPS09: A New Generation of NLO and LO Nuclear Parton Distribution Functions},
  \emph{JHEP} {\bf 0904} (2009) 065
  [arXiv:0902.4154 [hep-ph]].

\bibitem{Kovarik:2015cma}
  K.~Kovarik {\it et al.},
  \emph{nCTEQ15 - Global analysis of nuclear parton distributions with uncertainties in the CTEQ framework},
  arXiv:1509.00792 [hep-ph].


\bibitem{deFlorian:2011fp}
  D.~de Florian, R.~Sassot, P.~Zurita and M.~Stratmann,
  \emph{Global Analysis of Nuclear Parton Distributions},
  \emph{Phys.\ Rev.\ D} {\bf 85} (2012) 074028
  [arXiv:1112.6324 [hep-ph]].

\bibitem{Hirai:2007sx}
  M.~Hirai, S.~Kumano and T.-H.~Nagai,
  \emph{Determination of nuclear parton distribution functions and their uncertainties in next-to-leading order},
  \emph{Phys.\ Rev.\ C} {\bf 76} (2007) 065207
  [arXiv:0709.3038 [hep-ph]].

\bibitem{Paukkunen:2014slides}
  H.~Paukkunen, N.~Armesto and M.~Klein,
  \emph{Talk at LHeC Workshop 2014},
  \url{https://indico.cern.ch/event/278903/session/11/contribution/17}.

\bibitem{Paukkunen:2015slides}
  H.~Paukkunen, N.~Armesto and M.~Klein,
  \emph{Talk at LHeC Workshop 2015},
  \url{https://indico.cern.ch/event/356714/session/1/contribution/47}.

\bibitem{Khachatryan:2015hha}
  V.~Khachatryan {\it et al.} [CMS Collaboration],
  \emph{Study of W boson production in pPb collisions at $\sqrt{s_{\mathrm{NN}}}$ = 5.02 TeV},
  arXiv:1503.05825 [nucl-ex].

\bibitem{PaukkusenPuuhat}
  H.~Paukkunen, N.~Armesto and M.~Klein,
  \emph{work in progress}.

\bibitem{Albacete:2014fwa}
  J.~L.~Albacete and C.~Marquet,
  \emph{Gluon saturation and initial conditions for relativistic heavy ion collisions},
  \emph{Prog.\ Part.\ Nucl.\ Phys.}\ {\bf 76} (2014) 1--42,
  [arXiv:1401.4866 [hep-ph]].

\bibitem{Khachatryan:2015xaa}
  V.~Khachatryan {\it et al.} [CMS Collaboration],
  \emph{Nuclear effects on the transverse momentum spectra of charged particles in pPb collisions at $\sqrt{s_{_\mathrm {NN}}} =5.02$ TeV},
  \emph{Eur.\ Phys.\ J.\ C} {\bf 75} (2015) 5,  237
  [arXiv:1502.05387 [nucl-ex]].

\bibitem{ATLAS:2014cza}
  The ATLAS collaboration [ATLAS Collaboration],
  \emph{Charged hadron production in p+Pb collisions at $\sqrt{s_{_{\mathrm{NN}}}}=5.02$ TeV measured at high transverse momentum by the ATLAS experiment},
  ATLAS-CONF-2014-029, ATLAS-COM-CONF-2014-031.

\bibitem{Sassot:2009sh}
  R.~Sassot, M.~Stratmann and P.~Zurita,
  \emph{Fragmentations Functions in Nuclear Media},
  \emph{Phys.\ Rev.\ D} {\bf 81} (2010) 054001
  [arXiv:0912.1311 [hep-ph]].

\end{thebibliography}
\end{document}